\def\simless{\mathbin{\lower 3pt\hbox
           {$\rlap{\raise 5pt\hbox{$\char'074$}}\mathchar"7218$}}}    

\def\simmore{\mathbin{\lower 3pt\hbox
           {$\rlap{\raise 5pt\hbox{$\char'076$}}\mathchar"7218$}}}    

\documentclass[structabstract]{aa}

\usepackage{graphicx}

\begin{document}

\title{
Accretion and ejection in black-hole X-ray transients
}

\subtitle{}

\author{N. D. Kylafis\inst{1,2}
\and
T. M. Belloni\inst{3}
}

\institute{
University of Crete, Physics Department \& Institute of
Theoretical \& Computational Physics, 71003 Heraklion, Crete, Greece
\and
Foundation for Research and Technology-Hellas, 71110 Heraklion, Crete, Greece
\and
INAF-Osservatorio Astronomico di Brera, Via E. Bianchi 46, I-23807 Merate (LC), Italy
}

\date {Received ; Accepted ;}

\abstract
{
A rich phenomenology has been accumulated over the years regarding
accretion and ejection in black-hole X-ray transients (BHTs)
and it needs an interpretation.
}
{
Here we summarize the current observational picture of the outbursts of BHTs,
based on the evolution traced in a hardness - luminosity diagram (HLD), and
we offer a physical interpretation.
}
{
The basic ingredient in our interpretation is the 
Poynting - Robertson Cosmic Battery (PRCB, 
Contopoulos \& Kazanas 1998), which provides locally 
the poloidal magnetic field needed for the ejection of the jet.
In addition, we make two assumptions, easily justifiable.
The first is that the mass-accretion rate 
to the black hole in a BHT outburst has a generic bell-shaped
form, whose characteristic time scale is much longer than the dynamical
or the cooling ones.  This is guaranteed by the observational fact 
that all BHTs start their outburst and end it at the quiescent state, 
i.e., at very low accretion rate, and that state transitions 
take place over long time scales (hours to days).
The second assumption is that at low accretion rates the accretion flow is
geometrically thick, ADAF-like, while at high accretion rates it is 
geometrically thin.  
Last, but not least, we demonstrate that the previous history of the 
system is absolutely necessary for the interpretation of the HLD.
}
{
Both, at the beginning and the end of an outburst, 
the PRCB establishes a strong poloidal 
magnetic field in the ADAF-like part of the
accretion flow, and this explains naturally why a jet is always present in
the right part of the HLD.  In the left part of the HLD, the accretion
flow is in the form of a thin disk, and such a disk cannot sustain a 
strong poloidal magnetic filed.  Thus, no jet is expected in this part
of the HLD.  Finally, the counterclockwise traversal of the HLD is explained
as follows:  all outbursts start from the quiescent state, in which
the inner part of the accretion flow is ADAF-like, threaded by a poloidal
magnetic field.  As the accretion rate increases and the source moves
to the hard state, the poloidal magnetic field in the ADAF forces the flow
to remain ADAF and the source {{\it to move upwards in the HLD rather than
to turn left}}.  Thus, the {{\it history}} of the system determines the
counterclockwise traversal of the HLD.  As a result, no BHT is expected 
to ever traverse the entire HLD curve in the clockwise direction.
}
{
We offer a physical interpretation of accretion and ejection in BHTs 
with only one parameter, the mass transfer rate, plus the history of
the system.
}

\keywords{accretion, accretion disks -- X-ray binaries: black holes -- magnetic fields}

\authorrunning{Kylafis \& Belloni 2014}

\titlerunning{Accretion and ejection in black-hole X-ray transients}

\maketitle


\section{Introduction}

In the past decade, it has become clear that in order to characterise the 
spectral evolution of a black-hole X-ray transient (BHT) it helps to 
produce a hardness-luminosity diagram (HLD). Although separate outbursts 
even from the same source can look very different, the HLD shows a much 
clearer phenomenology, with most sources exhibiting a q-shaped curve, 
traveled counterclockwise, with additional excursions between soft and 
hard states (see Fig. 1, top panel; Homan et al. 2001; Belloni et al. 2005;
Homan \& Belloni 2005; Gierli\'nski \& Newton 2006; Remillard \& McClintock
2006; Fender et al. 2009; Motta et al. 2009; Belloni 2010; 
Munoz-Darias et al. 2011a; Stiele et al. 2011; Kylafis \& Belloni 2014). 
The fact that the diagram closes only at its end (lower right) indicates 
the presence of hysteresis: the source follows a different path in going 
from A to C than it does in going back from C to A.
The presence of hysteresis in the outburst of BHTs was already noticed 
in the pioneering work by Miyamoto et al. (1995).
Adding fast time variability to the picture in the form of rms integrated 
over a broad range of frequencies, two more useful diagrams have been 
introduced: the hardness - fractional-rms-variability diagram (HRD) 
(Fig. 1, bottom panel; Belloni et al. 2005)
and the absolute-rms-variability - intensity diagram (RID)  (Fig. 2; 
Munoz-Darias et al. 2011a,b).
Interestingly, looking at Figs. 1 and 2 one can see that two diagrams 
show clear hysteresis, but the HRD does not: the hard to soft branch 
overlaps the soft to hard one.
From these diagrams, together with a more detailed view of the fast time 
variability in terms of Power Density Spectra (PDS) a set of well defined 
states has been defined (Belloni et al. 2005; Belloni 2010; Belloni et al. 
2011). These are the quiescent
state (QS), the hard state, for historical reasons called low/hard
state (LHS),  the hard intermediate state (HIMS), the soft intermediate 
state (SIMS), and the soft state,
historically called high/soft state (HSS). The reason for the high/low in 
the state names is due to the fact that in the early observations, mostly 
of Cygnus X-1, the HSS was systematically brighter than the LHS at energies 
below 10 keV. However, at higher energies the situation is reversed. As we 
will show below, the idea that the HSS is observed systematically at higher 
accretion rate than the LHS persists, despite the fact that the observations 
of the last decade have shown otherwise.
A different classification scheme based on precise measurements of spectral 
and timing properties was introduced by Remillard \& McClintock (2006) and 
revised in McClintock et al. (2009). A comparison between the two 
classification schemes has been provided by Motta et al. (2009). 
For the purposes of this paper, we will use the first classification.

\begin{figure}
\centering
\includegraphics[angle=0,width=10cm]{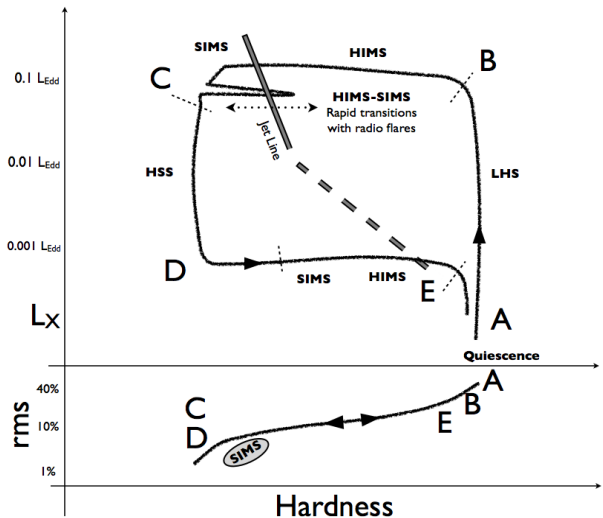}
\caption{
Schematic representation of the q-shaped curve in a HLD (top panel) 
and HRD (bottom panel) for black-hole X-ray binaries. The dotted lines 
in the top panel indicate the transition between states 
(marked with their acronyms). The arrows give the direction of motion 
along the line. The jet line is shown as a grey line in the HLD, 
dashed in the region where it might extend. The letters 
A, B, C, D, and E indicate turning points described in the text.
} 
\label{Fig1}
\end{figure}

\begin{figure}
\centering
\includegraphics[angle=0,width=10cm]{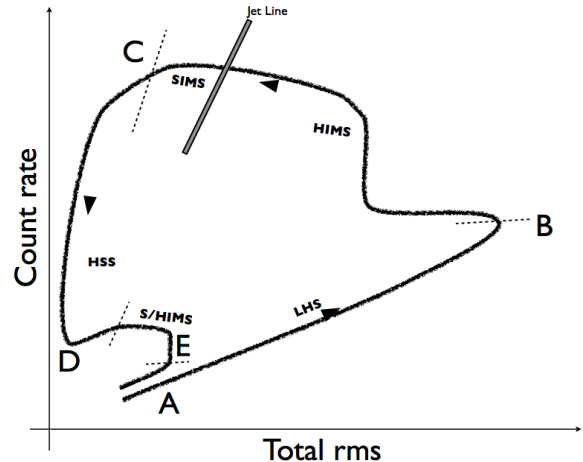}
\caption{
Schematic representation of the curve traversed by black-hole X-ray 
binaries in a RID. The dotted lines indicate the transition between 
states  (marked with their acronyms). The arrows give the direction of 
motion along the line. The jet line is shown as a grey line. 
The points A, B, C, D, and E mark the same points as in Fig. 1.
}
\label{Fig2}
\end{figure}

In a simplified and generic model, it is assumed that for all states 
during an outburst the accretion flow is composed of a cool (maximum 
temperature $\simless 1$ keV) geometrically thin and optically thick 
accretion disk as described by Shakura 
\& Sunyaev (1973), plus a hot corona near the black hole (the so-called 
``sombrero'' configuration, see Gilfanov 2010). The spectral state is 
determined by the relative size of the two physical components. 
In the LHS the corona is large, pushing the accretion disk farther out, 
in the HSS the disc extends all the way to the innermost stable  circular 
orbit (ISCO) and the corona is correspondingly much smaller in size.

A more realistic model was introduced by Esin et al. (1997; see also 
Narayan et al. 1996; Narayan et al. 1997; Esin et al. 1998, 2001), where 
the generic corona was replaced by a more physical Advection Dominated 
Accretion flow (ADAF; Narayan \& Yi 1994, 1995; Abramowicz et al. 1995). 
In our interpretation of the q-shaped curve, we will start from the 
Esin et al. (1997) model and will expand it to take into account the 
wealth of new information that has become available since then.

One important aspect of the q-path in the HLD beyond the presence of 
hysteresis, which is usually not considered, is its time direction. 
As we said, the curve is traversed counterclockwise: what prevents the 
opposite case, a clockwise path? In other words, why is the hard-to-soft 
transition always at a higher luminosity/accretion rate than the reverse 
transition (although both can be at different levels during different 
outbursts)? It is important to understand what breaks the time symmetry.
Not all sources exhibit a clean q-curve like the one shown in Fig. 1 
(based on the bright transient GX 339-4, see Belloni et al. 2005), 
but many do and we do not know a single case of a clockwise path. 
We will use GX 339-4 as the prototype.

For our interpretation, we make two basic assumptions: 

\begin{enumerate}

\item The accretion rate into the black hole has a generic bell-shaped curve 
as a function of time. By this we mean that the source will start from a 
very low accretion rate, increase steadily to a sizeable fraction of the 
Eddington rate, then decrease again down to a very low value. This is 
clearly justified by the fact that sources start and end their outbursts 
at very low luminosity (in the QS) and become brighter during the outburst. 
We do not mean that the accretion rate versus time can always be fitted 
with a Gaussian function.  It can certainly have local maxima and minima.
In addition, we assume that the characteristic time scale in this 
bell-shaped curve is much longer than the dynamical or 
cooling time scales.  This is justified by the fact that state transitions
take place over long timescales, typically days.
\item At high accretion rate, the accretion flow is geometrically thin and 
optically thick as in the solution of Shakura \& Sunyaev (1973), while at 
low accretion rate it is an ADAF (Narayan \& Yi 1994, 1995; Abramowitz et 
al. 1995) or ADAF-like.

\end{enumerate}
These two assumptions are commonly accepted and have been verified through 
numerical simulations (Ohsuga et al. 2009). 
 
Seen with the increased knowledge available now, the Esin et al. (1997) 
model has some shortcomings. 
In 1997, our picture was largely based on the results obtained 
on very few transients by the Ginga satellite (see Tanaka \& Lewin 1995). 
Their picture featured a Very High State (VHS) which appeared at accretion 
rates higher than those of the HSS, 
while we know now that the intermediate states 
(HIMS and SIMS) can appear over the same broad range of luminosities as 
the HSS. More important, their picture was ``one-dimensional'', in the 
sense that there was no hysteresis in the model.
The source states were sorted according to accretion rate: from QS to LHS 
to intermediate to HSS to VHS (see their Fig. 1, where the states are shown 
next to a vertical arrow indicating accretion rate). What is already clear 
by looking at the HLD is that {\it dynamically} this picture is correct: 
neglecting the complications introduced by intermediate states, the 
transition from hard to soft takes place as the accretion rate increases 
and the reverse transition as the accretion rate decreases. However, 
{\it statically} it is clear that you can have a HSS (point D in Fig. 1) 
at much lower accretion rate than  a LHS (point B in Fig. 1).
Finally, the presence of jets, their formation and their destruction were 
not known at the time and therefore were not part of the model.
Here, we extend the model of Esin et al. (1997) and its derivations 
(see Done et al. 2007 for a review) in order to account for the following:
\begin{itemize}
\item The return path (C-D-E in Fig. 1). The outburst decay has been 
studied in detail through several observational campaigns 
(see Kalemci et al. 2001, 2003, 2004, 2005, 2006), but to our knowledge 
it has not been included in theoretical models.
\item Jet formation, destruction, and re-formation. We will describe a 
novel mechanism for the generation of magnetic field in the ADAF part of 
the flow, which leads to an explanation for the three main observational 
facts: i) the formation and evolution of the compact jets observed in 
quiescence, hard, and hard intermediate states, ii) the eruptive disappearance 
of the jet once the {\it jet line} is crossed (roughly coincident with 
the HIMS-SIMS transition, see Fender et al. 2009), and iii) the 
gradual reappearance of the compact jet as the source reaches point E 
in Fig. 1. 
\end{itemize}
We will show that only one parameter, the mass accretion rate, plus
the history of the system, are 
sufficient to account for the full evolution of an outburst.

A crucial parameter, elusive to measure, is the transition radius $R_{tr}$ 
(Esin et al. 1997) between the outer geometrically thin thermal disk and 
the inner ADAF portion 
of the accretion flow. This radius is a function of accretion rate. 
Theoretical work on $R_{tr}$, which included evaporation, was presented by
Liu et al. (1999; see also Meyer et al. 2000; Meyer-Hofmeister \&
Meyer 2001; Qian et al. 2007; Meyer-Hofmeister et al. 2009).
Following this model, we show in Fig. 3 a sketch of the dependence of $R_{tr}$
with mass accretion rate $\dot M$. In the QS, $\dot M$ is small and 
$R_{tr}$ is large.
Between points A and B, in the LHS, $\dot M$ increases and $R_{tr}$ 
decreases. Between points
B and C, during the intermediate states, $\dot M$  continues to increase 
and $R_{tr}$ decreases down to 
$R_{\rm ISCO}$. Point C does not have to correspond to the maximum in 
$\dot M$, although it often does, but any 
further increase in accretion rate is not accompanied by a change in 
$R_{tr}$. In Fig. 3 we show three paths corresponding to three different 
accretion rate values at C. From point C the source is in the HSS. As 
the accretion rate starts to decrease,  $R_{tr}$ remains at $R_{\rm ISCO}$ 
until point D, when the HSS interval ends. At point D, the inner accretion disk 
becomes once again radiatively inefficient and increases its thickness 
(Das \& Sharma 2013). Approaching point E, $R_{tr}$ starts increasing and 
the inner ADAF region becomes larger until the loop closes and the source 
reaches once more the LHS.
Observationally, there is no agreement on the details of the variation of 
the transition radius as a function of 
accretion rate (see Sect. 2.3), but it is generally accepted that 
$R_{tr}$ is large in the QS and in the LHS, when 
$\dot M \simless 0.01 \dot M_E$, and is at its minimum in the HSS.

\begin{figure}
\centering
\includegraphics[angle=0,width=8cm]{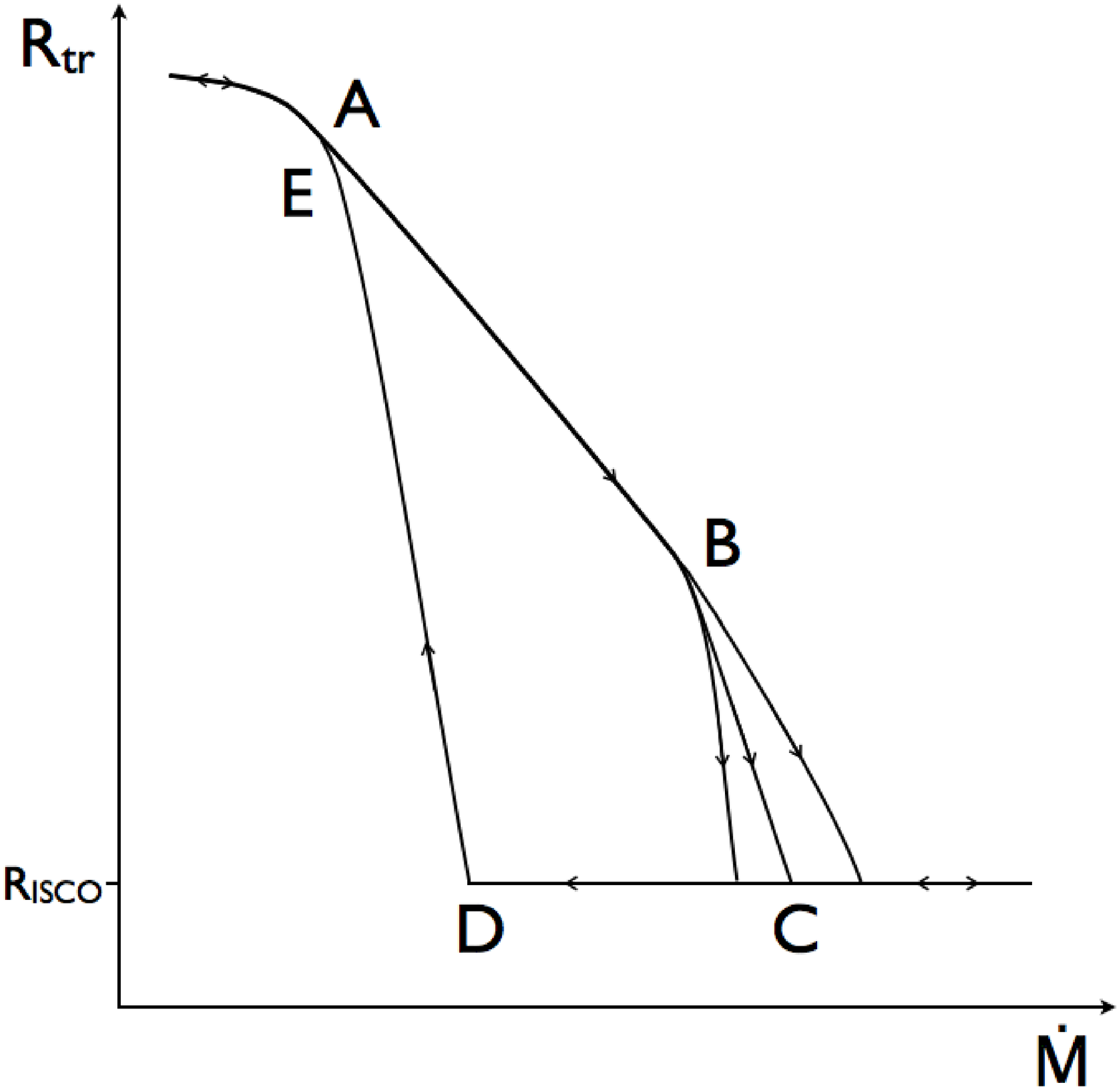}
\caption{
Schematic representation of the transition radius $R_{tr}$ between 
the inner ADAF and the outer thin disk as a function of mass-accretion 
rate $\dot M$. The points A, B, C, D, and E mark the same points as in 
Figs. 1 and 2. The arrows indicate the direction of motion along the lines.
}
\label{Fig3}
\end{figure}

The scheme we present in this work is of qualitative nature, but it 
relies on theoretical ideas and 
model calculations which are well accepted. Our description will be 
on the ``zeroth order''
effects. We will only offer partial speculations for additional details.
Our picture is developed for black holes, but many of its aspects can be 
applied to neutron-star and white-dwarf binaries
(Kylafis et al. 2012).

In \S~2 we outline the observational appearance of BHTs in the spectral 
and the timing domain,
in \S~3 we describe the general picture in order to model the 
``zeroth order'' phenomenology, in \S~4 we make some remarks,
and in \S~5 we present our conclusions.


\section{Spectral and timing properties of BHTs}

Here we present the spectral and timing properties of BHTs, which we will 
try to interpret below.

\subsection{X-ray spectrum}

The soft X-ray spectra (observed in the HSS) of BHTs are well 
interpreted with spectra of the 
form of a multi-temperature blackbody, modified by scattering and 
relativistic effects, and are
generally believed to originate from a geometrically thin and 
optically thick accretion disk of the type
proposed by Shakura and Sunyaev (1973). For a detailed analysis 
of the spectra see, e.g., Davis et al. (2005, 2006).

In the LHS, hard spectra are observed and there is much less 
agreement about their origin, although a general 
consensus is forming (see below). Models involving a corona 
(see Gilfanov 2010) posit that the spectra are
generated through Comptonization of soft seed (disk) photons 
by a population of thermal or non-thermal (or hybrid) 
electrons located in the corona. This class of models as such 
cannot account for the hysteresis described above.

All BHTs show radio emission that points toward evidence of 
the presence of a compact radio jet when they
are in their QS, LHS and HIMS (see, e.g., Fender \& 
Belloni 2012). Since in our picture the jet originates from 
an ADAF, which is associated with the generation of the 
observed hard spectra, it is natural to speculate that the jet
might play a role too. A simple jet model was shown to reproduce 
the X-ray spectra in the LHS and HIMS (Reig et al. 2003).
The same is true for ADAF models (see Done et al. 2007 for a review).
Given the presence of a jet and an ADAF, there is no need to introduce a 
corona, either thermal or non-thermal, to explain 
the LHS and HIMS hard spectra up to a few hundred keV. 
In fact, ADAF and jet models can 
address observational facts that are not included
in corona models (Giannios et al. 2004; Giannios 2005; Kylafis et al.
2008; Ingram et al. 2009; Ingram \& Done 2011, 2012). 
The upscattering of soft photons, which generates the hard spectrum, 
can take place at the base of the jet (Markoff et al. 2005).
However, in this scenario the photons will not be confined to the 
base of the jet, but will interact with the full compact jet. 
A model considering the full jet as scattering region fits the 
broad-band spectra from radio to X-rays (Giannios et al. 2004; 
Giannios 2005). 

In view of what has been outlined above, in this work we will assume 
that the hard spectrum observed in the LHS and HIMS is produced
by Comptonization of soft photons in the ADAF and possibly in 
the compact jet. The two possibilities are not that different, since 
the jet originates from the ADAF, which means that at ``zeroth order'' 
the mean electron energy in the ADAF and the jet must be comparable.

In the soft states, HSS and SIMS, in addition to the thermal disk, 
a steep power-law component is observed at high
energies, whose luminosity is only a small fraction of the total 
luminosity. This component has not received as much
attention. No jet emission from the core of the source has been 
observed in soft states, which points towards an alternative nature 
of this emission. A possible origin is through Comptonization in 
non-thermal flares above and below the thermal disk 
(Poutanen et al. 1997; Gierli\'nski et al. 1999). 

\subsection{Time variability}

Fast time variability is very state-dependent during outbursts of BHTs and 
it follows a very clear pattern (see Belloni 2010; Belloni et al. 2011).
Variations in time-variability properties can be followed in the diagrams
described above (Figs. 1 and 2).
The total fractional rms can be followed in the HRD
(bottom panel of Fig. 1). Here, as already remarked,
no hysteresis is visible: the source moves from a high amount of 
variability (30-40\%) in the LHS down to $\sim 1\%$ in the HSS
following a well defined path passing through the HIMS, and follows 
the same path on the return to quiescence. The only outliers 
correspond to the SIMS, which forms a separate ``cloud'' below 
the correlation.

The same evolution can be seen in the RLD (Fig. 2), for which 
no spectral information is needed. Hysteresis is observed in this 
diagram also. The SIMS is confined in a very small region 
of the RLD (see Mu\~noz-Darias et al. 2011a).


\section{Interpretation of the outbursts of BHTs}

Most of the BHT time is spent in the QS (see e.g. Levine et al. 2006), 
at a low accretion rate and luminosity. Only sensitive
instruments like those on board Chandra and XMM-Newton can detect 
them and for some of them long exposure times 
are needed (see, e.g., Gallo et al. 2006 and references therein).
For reasons which are thought to be connected to ionisation 
instabilities in the outer regions of the accretion disk 
(Meyer \& Meyer-Hofmeister 1981; Smak 1984; for a review see 
Lasota 2001), at some point a significant amount of matter 
starts accreting into the black hole and an outburst starts. 
Since all outbursts (with the exception of the peculiar 
GRS 1915+105, which started an outburst in 1992 and at the time 
or writing it is still active) terminate after weeks to months 
again to the QS, our ``zeroth order'' assumption of a bell-shaped 
time evolution of accretion rate is justified. Of course 
by ``bell-shaped'' we do not in any way mean that it has 
a Gaussian shape, but merely that it starts low, increases to 
a maximum or multiple local maxima, and then decreases again to the 
same value at which it started.  Our additional assumption,
that the characteristic time scale for changes in the accretion rate is
much larger than the dynamical and cooling time scales of the flow,
guarantees that the flow finds quickly its equlibrium.

Through multi wavelength observations, we know that at large 
distances from the black hole
($r \simmore 2 \times 10^4 R_g$, where $R_g=GM/c^2$ is the 
gravitational
radius and $M$ is the mass of the black hole) the accreting matter 
forms a thin disk
(McClintock et al. 2003; see also Marsh et al. 1994; Orosz et al. 1994).
As per our assumption, the inner region is in an ADAF state and 
accretion is at a very low level at the beginning of an outburst.

\subsection{Quiescent state}

We start in quiescence, well below point A in Fig. 1,
top panel. The accretion 
rate is very low and the accretion flow inside the 
transition radius $R_{tr}$ is well described by an ADAF. Radio 
emission associated to a compact jet
has been observed from BHTs in quiescence (see Gallo et al. 2006). The 
hard power-law energy spectrum is produced
by soft cyclotron photons from the jet upscattered by electrons 
either in the jet (Giannios 2005) or in the ADAF (Done et al. 2007).
A fraction of disk photons, coming from farther out in the accretion 
flow, are also scattered, but they contribute a small fraction of
the flux.

Three-dimensional magnetohydrodynamic simulations 
(Machida et al. 2006; Hawley 2009; Romanova et al. 2009;
Mignone et al. 2010) have shown that there are two mechanisms for 
jet formation, both requiring a strong large-scale magnetic field: 
plasma gun/magnetic tower (Contopoulos 1995; Lynden-Bell 1996), 
and centrifugal driving (Blandford \& Payne 1982).
Such a magnetic field can originate from a large distance and the 
advecting flow carries it to the inner region and amplifies it  
(Igumenshchev 2008; Lovelace et al. 2009; Tchekhovskoy et al. 2011).  
Alternatively, the field can be produced locally in the inner region, 
a possibility which we favor due to its simplicity and to the fact 
that no random process is involved.
A natural mechanism for the formation of a strong, poloidal magnetic 
field is the Poynting-Robertson Cosmic Battery (PRCB, 
Contopoulos \& Kazanas 1998; see also Contopoulos et al. 2006; 
Christodoulou et al. 2008). This mechanism is very efficient, because 
the inner region of the accretion flow is geometrically thick. 
This means that most of the radiation emitted near the ISCO will 
contribute to the battery mechanism. As for time scales 
(see Kylafis et al. 2012), since most of the life of BHTs is in 
the QS, there is more than sufficient time for the formation of 
the poloidal field needed for this mechanism to operate.

It is important to stress here the following:  the dynamical time
scale of the flow {\it is not} the fastest one in the problem.
The fastest time scale is the one for the diffusion of the magnetic 
field in the flow.  Thus, the magnetic field that is created by the 
Cosmic Battery diffuses outward in the flow {\it faster} than it is 
advected inwards (Contopoulos et al. 2006).  This allows its 
non-lenear growth.

\subsection{Hard state}

The efficiency of the Cosmic Battery is directly proportional to 
the luminosity. Therefore, as the source leaves the QS and enters 
the LHS, increasing its luminosity, the PRCB will produce a stronger 
magnetic field. The field will be able to support a stronger jet, 
brighter in the radio (e.g., Giannios 2005). This leads to a positive 
correlation between radio and X-ray luminosity (see Gallo 2010; 
Corbel et al. 2013a). We are now at point A in the HLD (Fig. 1).
As our source brightens, the inner accretion flow remains ADAF, 
the PRCB works more efficiently and the X-ray spectrum does not 
change dramatically in the 2-20 keV band, leading to a nearly vertical 
line in the HLD, with limited softening (see Motta et al. 2009).
However, the upscattering of more and more soft photons cools the 
electron population (Compton cooling; Done et al. 2007), leading to 
a decrease in the high-energy cutoff $E_c$ in the spectrum 
(Motta et al. 2009). The soft photon input increases because the 
transition radius moves inwards (Esin et al. 1997; Liu et al. 1999; 
Das \& Sharma 2013) and therefore the thermal component becomes 
hotter and more luminous. Above 1 keV, no direct emission is yet 
observable, so the only effect of the soft photons is to change 
the properties of the input photons for Comptonization 
(Sobolewska et al. 2011).

The ADAF inner flow within the transition radius is permeated by a 
strong poloidal magnetic filed, near equipartition, at the ISCO 
(Kylafis et al. 2012), which drops off with radius. This field 
supports the jet and increases the strength of the magneto-rotational 
instability (Bai \& Stone 2013). Because of this, there is
efficient transport of matter inwards and the density remains low. 
Thus, the source moves upwards in the q curve (towards point B,
Fig. 1, top panel) and does not turn left (towards point E).
This is where the {\it history} of the system plays a crucial 
role. From quiescence to this time, the inner part of the flow is
ADAF-like.  This ADAF is permiated by a strong poloidal magnetic 
field produced locally by the Cosmic Battery.  According to the work
of Bai \& Stone (2013), this magnetic field forces the source to move
upwards in the q-curve, i.e., to traverse it in the counterclockwise
direction.  Another consequence is that the upcoming state transitions
LHS-HIMS-SIMS-HSS (from point B to point C in Fig. 1) take place at 
higher accretion rate than in the return path (from point D to point E
in Fig. 1).  In the return path, the history is completely different
(see sections 3.6 - 3.10 below).  
From point C to point D in Fig. 1 the flow is in the form of a 
Shakura-Sunyaev thin disk and the Cosmic Battery is very inefficient in
producing a poloidal magnetic field.

Some authors suggested that in the LHS the transition radius is not 
located at several tens of gravitational radii, but it is close to the
ISCO (Miller et al. 2006a; Miller et al. 2006b). 
The small-radius scenario for the LHS has been challenged by 
Done et al. (2007). Recent works also appear to be inconclusive, 
with support for the small-radius picture (Reis et al. 2010; 
Reynolds \& Miller 2013) and for the opposite (Plant et al. 2013). 
We think that the truncated disk is to be preferred, as it 
explains physically the X-ray spectra and variability 
(Done et al. 2007; Ingram et al. 2009; Ingram \& Done 2011; 2012) 
and the jet formation/destruction (Kylafis et al. 2012).

In the LHS, 
the total variability is large, up to $\sim$40\% fractional rms. 
As the thin disk can vary only on a longer time scale, corresponding 
to the viscous time scale (Uttley et al. 2011, Wilkinson \& Uttley 2009), 
the observed variability must be generated in the hard component which 
dominates the flux, whether the ADAF or the jet. This was confirmed by 
Axelsson et al. (2013). As the source brightens and softens, the 
variability decreases and all characteristic frequencies 
increase, consistent with them being connected to decreasing radii.

In the HLD, moving from point A to point B along the LHS branch, 
the luminosity increases, the spectral photon index increases from 
$\sim$1.6 to $\sim$2.1, the high-energy cutoff decreases significantly 
(Motta et al. 2009) and, depending on accretion rate, interstellar 
absorption and observing X-ray band, the thin disk begins to contribute 
to the observed photons. At the top, the source turns left and enters the HIMS.

\subsection{Hard intermediate state}

At point B in the HLD (Fig. 1, top panel) the HIMS starts and the 
source moves left in the diagram, with a softening
spectrum. As a function of time, the accretion rate increases and 
$R_{tr}$ decreases. The ADAF section of the flow
becomes smaller and the thin disk extends closer to the black 
hole. Therefore, the relative contribution of the two 
parts of the accretion flow changes and the thin disk contributes more 
and more. In the energy band above $\sim$1 keV,
this is enhanced by the fact that the disk becomes hotter and 
contributes more photons to the detector.
As an example, Mu\~noz-Darias et al. (2011b) observe precisely 
this effect in MAXI J1659-152.

As the inner ADAF portion of the accretion flow becomes smaller, 
another effect causes the decrease in power-law
luminosity. Since the transition radius $R_{tr}$ becomes smaller, 
the jet becomes narrower, its optical depth lowers,
Comptonization becomes less efficient and the power law in the 
energy spectrum steepens (Kylafis et al 2008).
If the Comptonization takes place in the ADAF rather than in the jet, 
a similar argument can be made.
Moreover, as the geometrically thick ADAF shrinks to smaller 
radii, it will reach a point where the disk will not
be able to sustain the magnetic field produced by the PRCB 
(Kylafis et al. 2012). The jet becomes eruptive and is
made of discrete blobs with $\Gamma>2$ (Fender et al. 2004). When the 
so-called ``jet line'' is reached (Fender et al. 2004; Fig. 1),
the thin disk extends to the ISCO.

As the disk may support energetic flares (Poutanen et al. 1997; 
Gierli\'nski et al. 1999), the high-energy cutoff
$E_c$ is now determined by the non-thermal electrons in the flares. 
This explains the increase in $E_c$
observed in the HIMS as a function of time (see e.g. Motta et al. 2009). 
The high-energy spectrum is now of non-thermal 
nature (see also Grove et al. 1998).

In the timing domain, the PDS is the extension to higher frequencies 
of that in the LHS. The observed characteristic peaks  
correspond to those in the LHS, 
but with higher characteristic frequencies (e.g. Belloni et al. 2011). 
The characteristic frequencies move in unison (Belloni et al. 2005; 
notice that in Cyg X-1, a persistent system, one additional component 
was seen to remain at the same frequency across LHS-HIMS transitions, 
Pottschmidt et al. 2003). Therefore the system
displays effectively one frequency. The amount of variability, measured 
as integrated fractional variability, is lower than in the LHS (10-20\%). 
This can be interpreted within our picture. The thin disk, which shows 
little variability in this state (Gilfanov 2010), increases its 
contribution, lowering the total rms (Gierli\'nski \& Zdziarski 2005; 
Belloni et al., in preparation).
Since the variability originates from the ADAF/jet, higher variability 
is observed at higher energies (see, e.g., Gilfanov 
2010; Belloni et al. 2011). Notice that at high energies the spectrum 
is composed of two contributions: the ADAF and the energetic flares 
discussed above.

The truncated-disk model, with an inner ADAF and an outer 
Shakura-Sunyaev disk, explains the variability properties. 
A prediction of our picture is that in the upper branch of the q-diagram 
in the HLD (between points B and C in Fig. 1), the average hard time 
lag should {\it decrease} as the source softens, because as the ADAF 
region becomes smaller the travel time of the upscattered photons decreases.

\subsection{First jet-line crossing}

The jet line (Fig. 1, top panel) is defined as the position in the 
HLD corresponding to the ejection of a major transient relativistic
jet as observed in the radio band (Fender et al. 2004). 
Its location is close to that of the HIMS--SIMS transition (which 
is defined by marked and abrupt changes in the timing properties, 
see Belloni 2010), although an exact correspondence could not be 
made with the current data, rather sparse in radio (Fender et al. 2009). 
It also marks the disappearance of the compact jet 
(see Fender et al. 2004; Miller-Jones et al. 2012).
As discussed in Kylafis et al. (2012), at this stage the geometrically 
thin disk is not able to sustain the magnetic field produced by the 
PRCB and becomes unstable to non-axisymmetric 
``Rayleigh-Taylor-type'' instability modes.  The accumulated magnetic 
field escapes to the outer disk in the form of magnetic ``strands''
and this may explain the flaring activity.
While the timing properties change radically 
across the HIMS--SIMS transition, the energy spectrum below 10-20 keV 
changes only minimally, as recorded by the small change in hardness. 
A change in the evolution of the high-energy cutoff has however been 
observed (Motta et al. 2009).
The fast variability is stronger at high energies, which makes 
a disk origin unlikely.  In our picture, the variability is associated  
to the invoked magnetic flaring activity above and below the disk.

\subsection{Soft intermediate state}

The HIMS and the SIMS correspond to what in the pre-RXTE era was 
called Very High State (VHS), and Intermediate
State (see Miyamoto et al. 1993; Belloni et al. 1997). 
Now we know that as the accretion rate increases the sequence of 
states is LHS-HIMS-SIMS-HSS, but the HLD in Fig. 1, where the $y$ 
axis can be taken as a proxy for accretion rate, shows that the 
HSS does not correspond necessarily to a higher accretion rate than 
the LHS. The coverage with {\it Ginga} was not sufficient to show 
this sequence and affected the model of Esin et al. (1997).

In the SIMS we are to the left of the jet line. The state is defined 
by changes in the PDS. The identifying feature is the presence of the 
so-called type-B
QPO, which has been shown not to be an evolution of the one observed 
in the HIMS (type-C QPO; see Belloni et al. 2005; Casella et al. 
2004,2005; Motta et al. 2011,2012). 
Notice that the SIMS in the HRD is characterised by a drop in total 
fractional rms (the ellipse in Fig. 1, bottom panel). 
This is mainly due to the disappearing of the HIMS broad-band 
components in the PDS, which are replaced by a weaker power-law noise.

\subsection{Soft state}

After the SIMS is crossed, the source enters the HSS (point C in the 
HLD, Fig. 1). In outbursts like those of the prototypical 
source GX 339-4, where the HLD is similar to that in Fig. 1, the HSS 
after the SIMS marks the maximum accretion rate. From point C on, the 
accretion rate starts decreasing. However, there are sources where the 
accretion rate continues to rise even after the HSS is reached. 
They enter the ``anomalous state'' (Belloni 2010). We do not discuss 
this state because we consider it a ``first-order effect''. It is 
important to notice however that the anomalous state does not correspond 
to the VHS of Esin et al. (1997).

In the HSS, the energy spectrum is dominated by a soft thermal 
component, with the addition of an energetically negligible steep
power-law component with $\Gamma > 2.4$, which has been observed to 
extend up to 1 MeV (Grove et al. 1998). In all models, the thermal 
component is associated to a geometrically thin accretion disk 
extending all the way to the ISCO, while the hard power law could 
originate from Comptonization of soft photons by a non-thermal 
population of electrons in flares above and below the disk. 
As the mass accretion rate decreases, the accretion disk remains 
geometrically thin and extends to the ISCO. Since the luminosity 
decreases, the inner disk temperature also decreases and the hardness 
decreases (see Motta et al. 2010; Stiele et al. 2012). The source 
reaches point D.

The low level (a few \%) of fast variability observed in the PDS of 
the HSS comes from the power-law component and not from the thermal 
disk, as shown by Churazov et al. (2001), through the analysis of 
Fourier-resolved energy spectra.

\subsection{Soft intermediate state at low luminosity}

At point D, as the mass accretion rate continues to decrease, the inner 
portion of the disk becomes again
radiatively inefficient and thus geometrically thick (Das \& 
Sharma 2013), with evaporation 
playing a significant role (Liu et al. 1999). Simulations have 
shown that the transition radius between the two parts 
of the flow (the inner ADAF and the outer thin disk) moves outward 
as the mass accretion rate decreases
further (Das \& Sharma 2013).
Therefore, for the lower branch of the q-diagram (point D to point E), 
when the source hardens again showing
limited flux decrease, we have the reverse of what happened for 
the upper branch (point B to point C), with the
transition radius $R_{tr}$ moving outwards with time (see Fig. 3).

We remark here that, as the accretion rate decreases, 
evaporation of the inner part of the accretion disk and the initiation 
of an inner ADAF will take place at a more or less specific accretion rate.
This implies that the source will {{\it always}} turn right at about 
point D (i.e., at a specific accretion rate), 
independent of how high the luminosity was in the
upper branch of the q-shaped curve (points B and C in Fig. 1,
top panel).  This is exactly what has been seen in the multiple 
outbursts of GX 339-4 (Motta et al. 2011).
With the presence of a geometrically thick ADAF in the inner part of 
the flow, the PRCB can again operate efficiently
in creating the magnetic field necessary for the production of a jet. 
The SIMS is entered again, as indicated by
the appearance of type-B QPOs (Stiele et al. 2011; Motta et al. 2011).

A radio compact jet is observed only later, when the source has reached 
the hard state, but its formation must have started
earlier. A gradual development of the jet has been observed in detail 
(Corbel et al. 2013b and references therein). Notice that the
photon index of the hard component in the SIMS in the lower branch 
($\sim$2.1, see Stiele et al. 2011) is lower than that  
of the upper branch ($\sim$2.4).

\subsection{Hard intermediate state at low luminosity}

The SIMS-HIMS transition takes place on the lower branch as the 
accretion rate decreases and the source hardens. Corresponding
to this transition, no eruptive jet similar to those in the upper 
branch has been observed. We predict that none will be observed, 
because as the ADAF inner portion of the flow continues to increase, 
the PRCB will continue working and the jet will build up smoothly. 
In the lower branch, no instability occurs, although as the jet is 
formed variability is naturally expected.

The spectral evolution is symmetric to what took place in the upper 
branch. The ADAF flow inside $R_{tr}$
adds to the energy spectrum a hard component, deriving from 
upscattering of softer disk photons in the inner flow or
in the forming jet. The optically thick disk part recedes and its 
temperature decreases, also leading to a hardening of the
energy spectrum. This is what was observed by Munoz-Darias et al. 
(2011b) for MAXI J1659-152.

In our picture, we make the prediction that in the lower branch of 
the q-shaped curve in the HLD the average hard time lag should
{\it increase} as the source hardens. This is because the ADAF region, 
where the Comptonization takes place, increases its radius
(Das \& Sharma 2013) and therefore the light travel time for 
Comptonized photons increases.

\subsection{Second jet-line crossing}

The source is now around point E in the HLD (Fig. 1). As we discussed 
above, the jet line corresponding to the HIMS-SIMS
transition does not have a counterpart at the reverse SIMS-HIMS 
transition. We define the crossing of the jet line, 
or its extension at low fluxes, as the time at which the compact radio 
jet is observed again. This occurs shortly before the full hard state 
is reached (Miller-Jones et al. 2012).
At the time of this reverse crossing, a large part of the inner 
accretion flow is geometrically thick and $R_{tr}$ is
now at tens to hundreds gravitational radii and the energy spectrum 
is harder than at the first jet line crossing. This leads to a diagonal
jet line in the HLD in Fig. 1.

\subsection{Back to the hard state}

The LHS is reached again when the ADAF flow takes most of the X-ray 
emitting region of the disk. As in the case of the 
start of the outburst, the seed photons for Comptonization are 
mainly cyclotron photons from the jet and the Comptonization
itself takes place either in the jet itself or in the ADAF, with 
a small contribution from the disk.

\subsection{Return to the quiescent state}

The accretion rate continues to decrease and the reverse of the 
initial LHS takes place, with the magnetic field produced by the 
PRCB becoming weaker, making the jet fainter, with the energy 
spectrum hardening only marginally. The source goes to quiescence 
remaining in the LHS and leaving the HLD in Fig. 1 from the bottom 
side along the same path as at the start.

\section{Remarks}

The curve shown schematically in the HLD (Fig 1. top panel) is a 
hysteresis curve, as it is observed in many branches
of science. All such curves represent cases when the future development 
of a system depends on its past history in addition to its present.
In the case of black hole binaries, the poloidal magnetic field created 
by the PRCB made possible by the ADAF in the quiescent and hard states 
``reminds'' the flow that it must remain in an ADAF state, although 
the accretion rate is increasing. Only at high accretion rate
the thermal disk prevails and there is a state transition.

Therefore, the stability of the soft and hard states at the same 
luminosity level can be described as follows.
Let us take an accretion rate such that the observed 
luminosity puts the source between the upper and lower
branches in the HLD. For this accretion rate, two states are possible: 
LHS on the right branch (ADAF-like) or HSS (Shakura-Sunyaev-like) 
on the left branch. This is a {\it static} view: given only the 
accretion rate, we cannot state where the source could be found. 
However, when moving from A to B in Fig. 1, the poloidal magnetic 
field that necessarily exists at $R_{tr}$
``reminds'' the accretion flow that it should retain 
its flow efficiency (Bai \& Stone 2013), and therefore remain ADAF, 
{\it despite the increase of the accretion rate}.
Thus, the source moves vertically from point A to point B and does not
turn left.  Similarly, when the source moves from C to D in Fig. 1,
radiative efficiency in the disk forces the disk to remain thin.

In summary, the history of the system, in addition to a single 
parameter (accretion rate) are sufficient to explain qualitatively 
the hysteresis  curve.
We remark that no astrophysical models can explain all observational 
details and our proposed zeroth-order picture is no exception. Future 
work will address ``first-order'' effects, such as the presence and 
nature of the anomalous state (Belloni 2010), the different levels 
of transition luminosity for multiple outbursts of GX 339-4 (where 
the q shapes are ``nested'' within each other, see e.g. Belloni 2010), 
the repeated HIMS-SIMS-HIMS transitions with multiple crossings of 
the jet line, the sources which do not display such a well behaved 
q diagram, and so on.
			
\section{Conclusions}

Making use of a) the novel idea of the Poynting-Robertson Cosmic
Battery (Contopoulos \& Kazanas 1998), b) two assumptions easily 
justifiable, and c) the history of BHTs, we have been able  
to interpret the major observational properties of
the evolution of outbursts of BHTs. The assumptions are:

\begin{enumerate}

\item The accretion rate onto the central object has a generic 
bell-shaped curve, starting from a very low level,
increasing steadily up to a level that can be a sizeable fraction 
of the Eddington value, and decreasing again to a very low level.
This is justified by the fact that BHTs start and end their outbursts 
in quiescence, at very low luminosity and that during their outburst 
they reach a high luminosity, often close to Eddington.
In addition, we assume that the time scale for accretion-rate changes
is much longer than the dynamical and cooling time scales.

\item At high accretion rates, the accretion flow solution is that of 
an optically thick and geometrically thin disk (Shakura \& Sunyaev 1973).
At low accretion rates, it is radiatively inefficient, geometrically 
thick, optically thin and advection dominated (ADAF; Narayan \& Yi 
1994, 1995; Abramowitz et al. 1995). Both these pictures are widely 
accepted and have been tested with numerical simulations 
(Ohsuga et al. 2009).
\end{enumerate}

When the accretion flow possesses an inner ADAF portion, the PRCB 
works efficiently (Kylafis et al. 2012). Therefore, a strong
poloidal magnetic field is formed and a jet can be built on time scales 
of hours to days, depending on the luminosity. This is what is observed 
in the HLD: on the right track, the accretion rate varies considerably, 
a large part of the flow is ADAF-like and a compact jet is present.

The sourses traverse the HLD in the counterclockwise direction,
because their history plays a significant role.  All 
outbursts start at low luminosity, when the sources are in the quiescent
state.  In this state, due to the low accretion rate, the major part
of the flow, its inner part, is ADAF-like and the PRCB works efficiently.
The higher the accretion rate, the higher the luminosity, the larger
the efficiency of the PRCB, and the stronger the poloidal magnetic field
that is produced there.  Thus, the sources are {{\it forced}}
(Bai \& Stone 2013)
to move upwards in the HLD (counterclockwise direction) rather than move 
to the left (clockwise direction).

At high accretion rates, the flow switches to a full Shakura-Sunyaev 
solution. In this case, the PRCB is no longer efficient and 
is not able to form a strong poloidal magnetic field to keep producing 
a jet. No jet has been observed from a source in the soft 
state. 

In this picture, it is natural to expect that a {\it transient jet} 
is produced in the first jet line crossing (hard to soft) but not in 
the second (soft to hard). As the first jet line is approached, the ADAF 
part of the flow, the one which feeds the jet, shrinks. The 
geometrically thin disk cannot sustain the existing field and becomes 
unstable to instability modes of Rayleigh-Taylor type. The accumulated
magnetic field then escapes and a flare is produced by magnetic reconnection.
During the second jet line crossing, the low luminosity implies that the 
poloidal magnetic field is built slowly in the ADAF and grows
steadily. No transient ejection can be produced in this case.

In summary, we have shown that the main observed features along 
the q-shaped curve in the HLD can be reproduced with only one
physical parameter, the mass accretion rate. 
Additional ingredients in our picture are the Poynting-Robertson
Cosmic Battery (Contopoulos \& Kazanas 1998) and the history of the system.
The q-shaped curve 
will always be traversed in the counterclockwise direction and our 
prediction is that no 
clockwise evolution will be seen in the future. In addition, we 
predict that the average hard time lags decrease with time along the 
upper branch and increase with time along the lower branch.
For the formation and the destruction of the jet, our picture can be 
applied also to neutron-star and white-dwarf binaries (Kylafis et al. 2012).

%
\begin{acknowledgement}
We thank an anonymous referee for helpful suggestions and comments, 
which have improved our paper in both content and readability.
We have also profited from discussions
with Iossif Papadakis.
One of us (NDK) acknowledges useful discussions
with P. Casella, I. Contopoulos,
B. F. Liu, S. Motta, R. Narayan, and A. Zdziarski.
This research has been supported in part by 
the ``RoboPol" project, which is implemented under the ``ARISTEIA" 
Action of the ``OPERATIONAL PROGRAM EDUCATION AND LIFELONG LEARNING" 
and is co-funded by the European Social Fund (ESF) and National Resources.
Also, by EU Marie Curie projects
no. 39965 and ITN 215212 (``Black Hole Universe''), EU REGPOT project 
number 206469, a Small Research Grant from the University of Crete,
a COST-STSM-MP0905 grant and a grant from the European Astronomical Society
in 2012.  TMB acknowledges support from INAF-PRIN 2012-6.
\end{acknowledgement}

\end{document}